\documentstyle[11pt,newpasp,twoside,epsf]{article}
\def\edcomment#1{\iffalse\marginpar{\raggedright\sl#1\/}\else\relax\fi}
\reversemarginpar

\begin{document}
\title{The Brown Dwarf Deficit in Taurus: Evidence for a Non-Universal IMF}
 \author{C\'esar Brice\~no}
\affil{Centro de Investigaciones de Astronom{\'\i}a (CIDA), 
Apartado Postal 264, M\'erida 5101-A, Venezuela}
\author{Kevin Luhman \& Lee Hartmann}
\affil{Harvard-Smithsonian Center for Astrophysics, 60 Garden St., Cambridge, MA 02138, USA}
\author{John Stauffer \& Davy Kirkpatrick}
\affil{SIRTF Science Center, MS 220-6, Caltech, Pasadena, CA 91125, USA}
\author{Davy 0}
\affil{IPAC, MS 100-22, 770 S. Wilson, Caltech, Pasadena, CA 91125, USA}

\begin{abstract}
We present the results of a deep, optical/IR wide field imaging survey of 
selected fields in the nearby (d$\sim 140$ pc) Taurus star-forming region.
We report the discovery of 9 new members with spectral types M5.75-M9.5.
We derive an Initial Mass Function encompassing 54\% of the known members in
Taurus. Comparison with dense regions like the Trapezium Cluster in Orion shows
that Taurus has produced $\times 2$ less brown dwarfs. 
We suggest that the lower
frequency of brown dwarfs in Taurus may result from the low-density
star-forming environment, leading to larger minimum Jeans masses.
\end{abstract}

\section{Introduction}

The variation of the Initial Mass Function (IMF) with the 
environment is a key problem in star formation,
but any dependence of the IMF with the ambient conditions 
has been difficult to determine. Most
surveys have concentrated on the most compact regions with the 
highest density of stars and gas, while extended, sparse groups of 
nearby young stars have not been studied with the same sensitivity, 
mainly because of the large area of the sky that they span. 
Therefore, meaningful comparisons of the IMF in
differing conditions have not been readily available.
This situation is changing 
with the advent of large-format CCD detectors, making possible 
deep surveys of the less embeded, more extended star-forming complexes.
With a modest extinction (Av $\la 4$) and low density of stars
(N$\sim 1-10$ pc$^{-3}$) Taurus is an example of the ``non-clustered'' mode
of star formation, making it an ideal laboratory for comparison 
with high density regions like young clusters.

\section{Observations and Results}

 We used the 8k $\times $8k CCD Mosaic Camera with $I,z'$ filters
on the KPNO 0.9m telescope ($1^o \times 1^o$ FOV)
to observe eight fields in Taurus to an $I_{lim}\sim 21$. 
With $I,z'$ photometry and $JHK_s$ data from the 2MASS, we performed
a reddening limited (Av$\la 4$) search for candidate young ($\la 10$ Myr)
low-mass stars and brown dwarfs down to $0.02\> {\rm M_{\odot}}$
(details in Brice\~no et al. 2002).
We then used the K and Na gravity-sensitive 
features in follow up spectra to confirm membership.

Among the new members is an M9.5 brown dwarf (Figure 1a), the
least massive object identified so far in Taurus ($15\> {\rm M_{Jup}}$
using the temperature scale of Luhman [2000] and evolutionary tracks 
of Baraffe et al. [1998]).  This object exhibits
significant near-IR excess and strong H$\alpha$ emission
(W[H$\alpha]=150$\AA), indicative of ongoing accretion
from a circumstellar disk and evidence that brown dwarfs, like stars,
can form from fragmentation processes.

We construct an IMF for our Taurus fields and compare it with the IMF
for the Trapezium Cluster (Figure 1b). 
We find that the ratio of substellar to stellar
objects in the Trapezium is $\times 2$ larger than in Taurus;
the scarcity of brown
dwarfs may be due to the higher value of the Jeans mass in Taurus,
which results from the lower surface density of gas in this region.

\begin{figure}
\plottwo{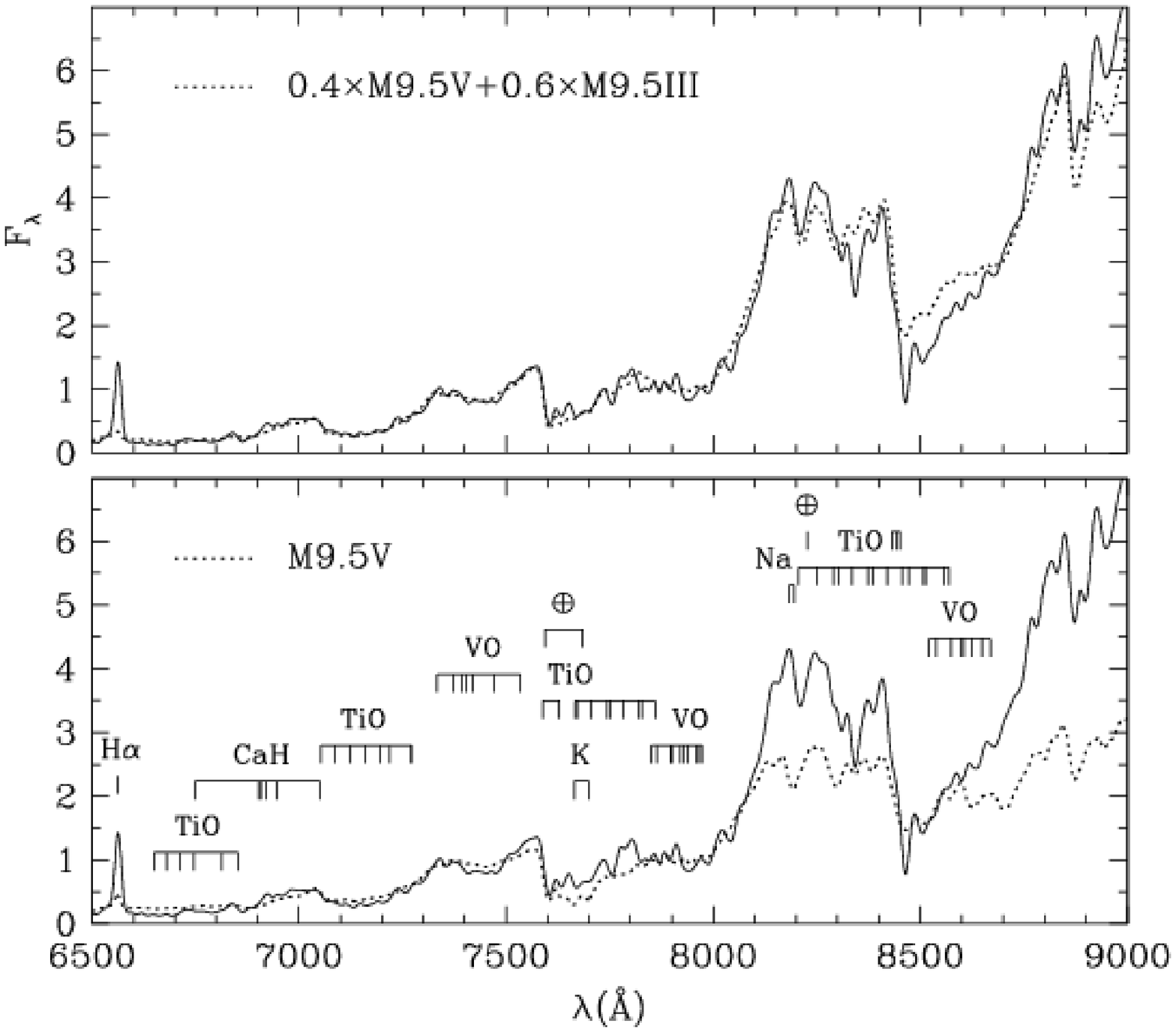}{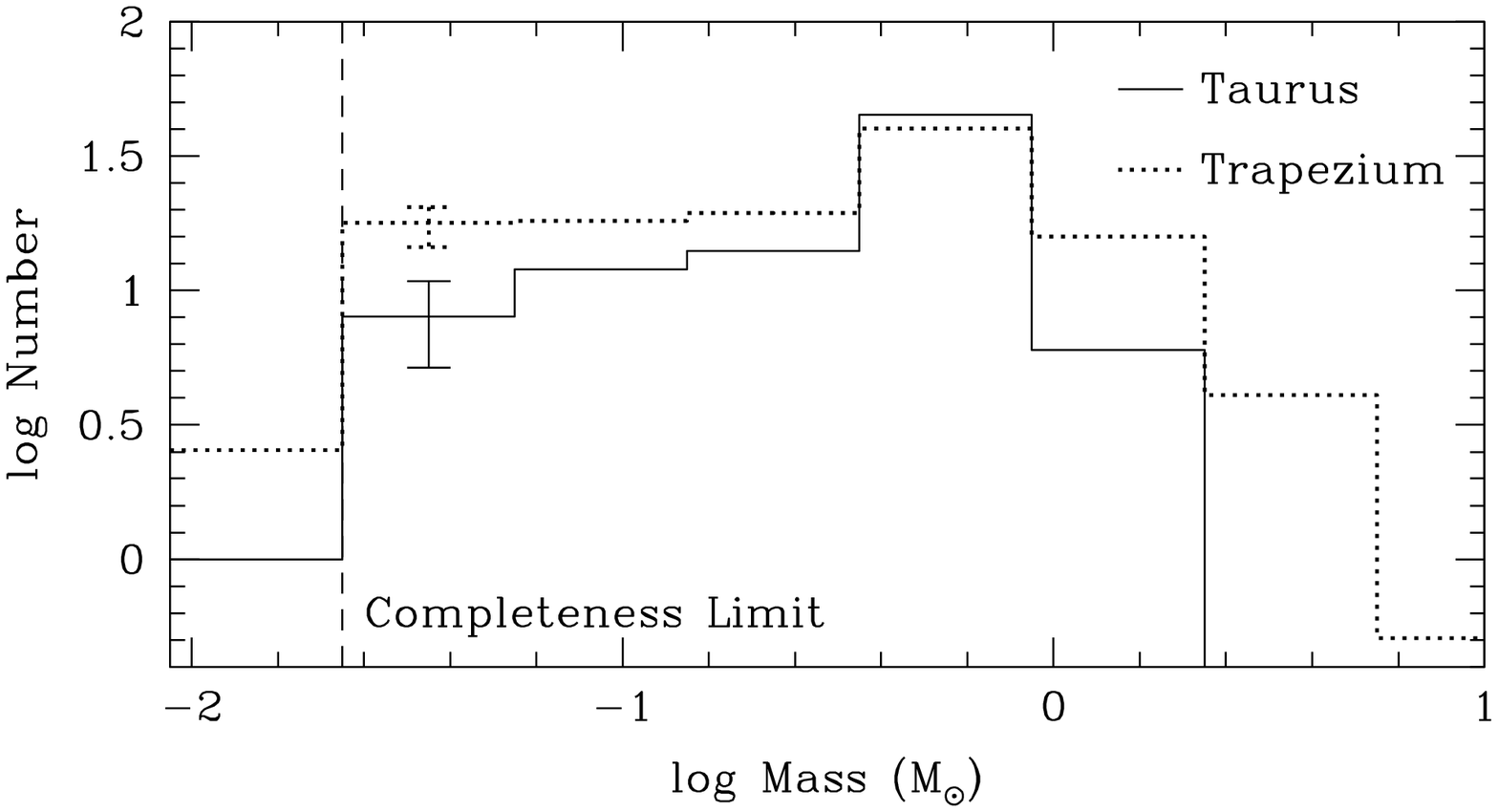}
\caption{(a) Spectra for the M9.5 brown dwarf. The best fit spectrum 
(upper panel) is obtained for an intermediate gravity between dwarfs 
and giants.
(b) IMFs for reddening limited samples in Taurus (solid line) 
and the Trapezium Cluster (dotted line). 
The Taurus sample was normalized to the Trapezium.}
\end{figure}

\end{document}